\documentclass{emulateapj}

\begin{document}

\shortauthors{Luhman et al.}
\shorttitle{One of the Coldest Brown Dwarfs}

\title{Confirmation of One of the Coldest Known Brown Dwarfs\altaffilmark{1}}

\author{
K. L. Luhman\altaffilmark{2,3},
A. J. Burgasser\altaffilmark{4},
I. Labb\'e\altaffilmark{5},
D. Saumon\altaffilmark{6},
M. S. Marley\altaffilmark{7},
J. J. Bochanski\altaffilmark{2},
A. J. Monson\altaffilmark{5},
and S. E. Persson\altaffilmark{5}
}

\altaffiltext{1}{Based on observations made with the following facilities:
the Spitzer Space Telescope, which is operated by the Jet Propulsion
Laboratory, California Institute of Technology under a contract with NASA;
the ESO Telescopes at Paranal Observatory under program ID 286.C-5042;
the 6.5 meter Magellan Telescopes located at Las Campanas Observatory,
Chile.}
\altaffiltext{2}{Department of Astronomy and Astrophysics, The Pennsylvania
State University, University Park, PA 16802, USA; kluhman@astro.psu.edu}
\altaffiltext{3}{Center for Exoplanets and Habitable Worlds, The 
Pennsylvania State University, University Park, PA 16802, USA}
\altaffiltext{4}{Center for Astrophysics and Space Science, University of
California San Diego, La Jolla, CA 92093, USA}
\altaffiltext{5}{Observatories of the Carnegie Institution for Science,
813 Santa Barbara St., Pasadena, CA 91101, USA}
\altaffiltext{6}{Los Alamos National Laboratory, P.O. Box 1663, MS F663,
Los Alamos, NM 87545, USA}
\altaffiltext{7}{Space Science and Astrobiology Division, NASA Ames Research
Center, Mail Stop 245-3, Moffett Field, CA 94035, USA}

\begin{abstract}

Using two epochs of 4.5~\micron\ images from the
Infrared Array Camera (IRAC) on board the {\it Spitzer Space Telescope},
we recently identified a common proper motion companion to the
white dwarf WD~0806-661
that is a candidate for the coldest known brown dwarf.
To verify its cool nature, we have obtained images of this object
at 3.6~\micron\ with IRAC, at $J$ with HAWK-I on the Very Large Telescope,
and in a filter covering the red half of $J$ with FourStar on Magellan.
WD~0806-661~B is detected by IRAC but not HAWK-I or FourStar.
From these data we measure colors of $[3.6]-[4.5]=2.77\pm0.16$ and
$J-[4.5]>7.0$ (SNR$<$3).
Based on these colors and its absolute magnitudes, WD~0806-661~B
is the coldest companion directly imaged outside of the solar system
and is a contender for the coldest known brown dwarf with the Y dwarf
WISEP J1828+2650. 
It is unclear which of these two objects is colder given the available data.
A comparison of its absolute magnitude at 4.5~\micron\ to the
predictions of theoretical spectra and evolutionary models
suggests that WD~0806-661~B has $T_{\rm eff}=300$--345~K.

\end{abstract}


\section{Introduction}
\label{sec:intro}

The coolest brown dwarfs are valuable laboratories for studies of the
atmospheres and interiors of substellar objects, including gas giant planets. 
Recent years have seen significant progress in extending samples
of brown dwarfs below 700~K ($\gtrsim$T8). Many of these objects
have been found at near-IR wavelengths through wide-field surveys
\citep{war07,del08,del10,burn08,burn09,burn10,luc10,alb11} and 
high-resolution imaging of substellar primaries \citep{liu11a,gel11}.
Because the coolest brown dwarfs are brightest at wavelengths
longward of 4~\micron\ \citep{bur03,sau08},
a sensitive mid-IR telescope is the best tool for detecting them.
One facility of this kind is the {\it Wide-field Infrared Survey Explorer}
\citep[{\it WISE},][]{wri10}, which obtained mid-IR images
of the entire sky in 2010. Data from {\it WISE} have already been used to 
dramatically increase the number of known late-T dwarfs
\citep{mai11,bur11,sch11,kir11,liu11b} and to discover the first objects that
show clear spectroscopic signatures of a new spectral class later than T
\citep[``Y dwarfs",][]{cus11}.

The {\it Spitzer Space Telescope} complements {\it WISE} by providing greater
sensitivity for smaller fields \citep{wer04}. As a result, 
{\it Spitzer} is well-suited for finding brown dwarfs as companions
to nearby stars \citep{luh07}.
To employ {\it Spitzer} for a companion survey, we have
searched multi-epoch {\it Spitzer} images for objects that are comoving
with stars in the solar neighborhood. Through this work, we recently identified
a common proper motion companion to the white dwarf WD~0806-661 \citep{luh11}.
The 4.5~\micron\ absolute magnitude of WD~0806-661~B suggests that
it could be the faintest and coolest known brown dwarf ($T_{\rm eff}\sim300$~K).
In addition to potentially breaking new ground in terms of temperature,
this companion could be one of a small number of benchmark brown dwarfs
whose ages are known via their primaries and thus can provide unusually
stringent tests of theoretical models of substellar objects.
However, the temperature of WD~0806-661~B is uncertain since it has
been detected at only 4.5~\micron.

To confirm its substellar nature and better constrain its temperature,
we have obtained deep near- and mid-IR images of WD~0806-661~B
with {\it Spitzer} and ground-based telescopes.
In this paper, we describe the collection and reduction of these data
(Section~\ref{sec:obs}), use the resulting astrometry and photometry
to assess the companionship and properties of WD~0806-661~B
(Section~\ref{sec:analysis}), and discuss the implications
of this object for theoretical models and surveys for cool brown dwarfs
(Section~\ref{sec:disc}).

\section{Observations}
\label{sec:obs}

\subsection{Spitzer IRAC}
\label{sec:irac}

WD~0806-661~B was imaged previously at 4.5~\micron\ by the
Infrared Array Camera \citep[IRAC;][]{faz04} on board {\it Spitzer}
in 2004 and 2009, as described by \citet{luh11}.
To measure a color relative to 4.5~\micron\ that is sensitive to temperature
for cool brown dwarfs, we elected to observe WD~0806-661~B at
3.6~\micron\ with IRAC \citep{pat06}.
This camera has a plate scale and a field of view of
$1\farcs2$~pixel$^{-1}$ and $5\farcm2\times5\farcm2$, respectively,
and produces images with FWHM$=1\farcs6$ at 3.6~\micron.
On 2011 April 18, IRAC collected one 93.6~s exposure of WD~0806-661~B
through the 3.6~\micron\ filter at each position in a 12-point dither
pattern.  These observations were conducted through Astronomical
Observation Request 41905408 for program 70203.

The Spitzer Science Center pipeline (version S18.18.0) 
performed the initial processing of the individual 3.6~\micron\ images.
We then combined these images into one mosaic using the Overlap and
Mosaic pipelines within the Mosaicking and Point-source Extraction
software package \citep[MOPEX,][]{mak05}.
We used MOPEX rather than the software applied to the 4.5~\micron\ data by
\citet{luh11} since it offers drizzling and Point Response Function (PRF)
subtraction, which are beneficial for our analysis of WD~0806-661~B, as
described below. We re-reduced the two sets of 4.5~\micron\ images with MOPEX as
well. A plate scale of $0\farcs6$~pixel$^{-1}$ was selected for the final
mosaics. To optimize the accuracy of the relative astrometry between the three
mosaics, we measured new world coordinate systems (WCSs) for the
first and third epochs of data based on the second epoch mosaic
in the manner described by \citet{luh11}.

\begin{figure}
\epsscale{0.7}
\plotone{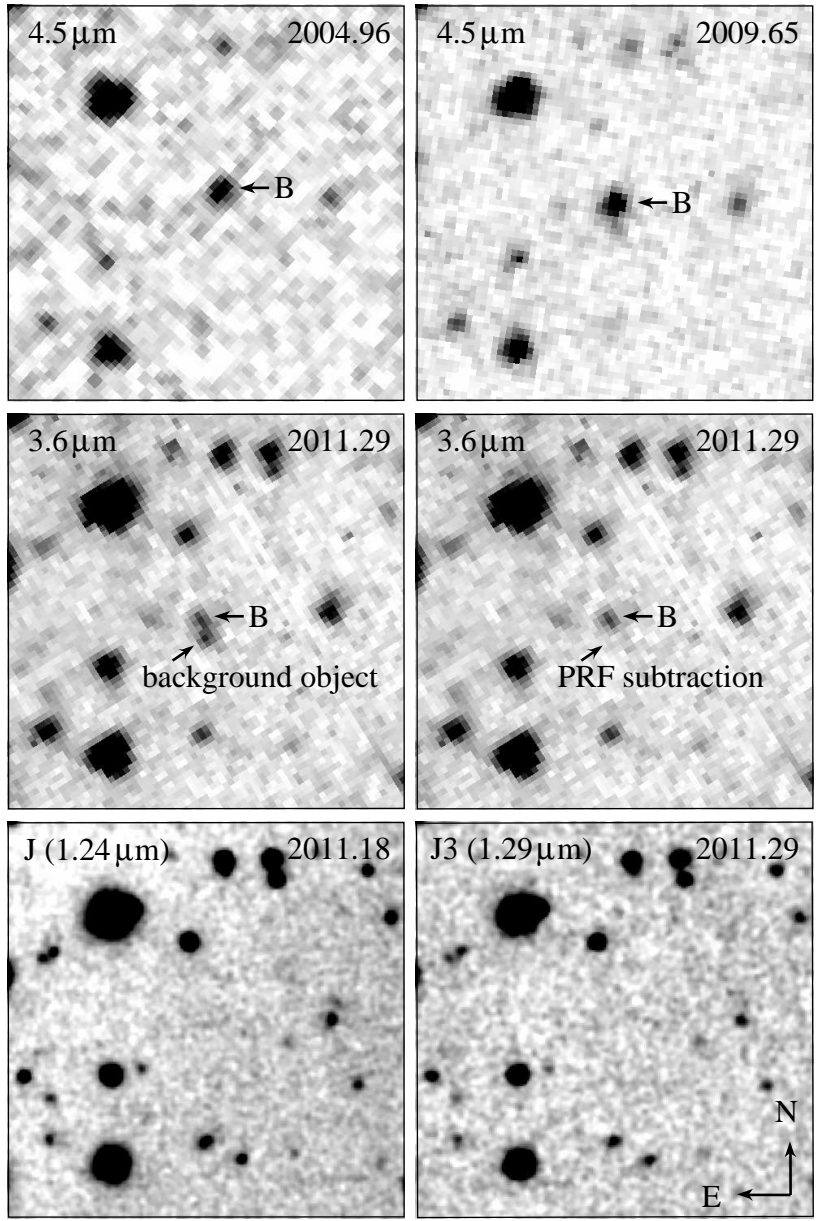}
\caption{
Discovery images of WD~0806-661~B at 4.5~\micron\ \citep{luh11} and
new data at 3.6~\micron, $J$ (1.24~\micron), and $J3$ (1.29~\micron).
WD~0806-661~B is detected at 3.6~\micron\ but not in $J$ and $J$3.
It is partially blended with a background object in
the 3.6~\micron\ image. We have applied PRF subtraction to the latter
prior to measuring astrometry and photometry for WD~0806-661~B.
The size of each image is $40\arcsec\times40\arcsec$.
}
\label{fig:image}
\end{figure}

The reduced IRAC images surrounding the position of WD~0806-661~B
are shown in Figure~\ref{fig:image}. We indicate the detections of
the companion in the first two epochs at 4.5~\micron. WD~0806-661~B
is detected in the new image at 3.6~\micron\ at the location expected
based on its astrometry and proper motion from the 4.5~\micron\
data (Section~\ref{sec:pm}), although it is much fainter than at 4.5~\micron.
In the 3.6~\micron\ image, WD~0806-661~B is partially blended with another
source at a distance of $\sim2\arcsec$ that is similar in brightness, which
is presumably a background star or galaxy.
The use of drizzling within MOPEX has optimized the spatial resolution
of the final mosaic, making it slightly easier to resolve WD~0806-661~B
from this source.
Through the Apex pipeline within MOPEX, we applied PRF fitting to the pair
of objects and subtracted the fit to the background source, as shown in
Figure~\ref{fig:image}.
The background object also is visible in the wing of WD~0806-661~B 
in the second 4.5~\micron\ image, but because WD~0806-661~B is much brighter
at that wavelength, we did not notice its presence in \citet{luh11}.
We have now attempted to subtract the contribution from this source in
this 4.5~\micron\ image in the same manner as done at 3.6~\micron.
The aperture for photometry of WD~0806-661~B in the first epoch at
4.5~\micron\ does not encompass the location of the background source.

We measured aperture photometry for WD~0806-661~B in each of the three
mosaics with an aperture radius of 4 pixels ($2\farcs4$) and radii of 5 and 10
pixels for the inner and outer boundaries of the sky annulus, respectively.
To estimate aperture corrections, we measured photometry for all sources
in a given mosaic with these parameters and with those used by \citet{luh10tau},
computed the average offset between the two sets of photometry, and combined
this offset with the aperture corrections measured by \citet{luh10tau}.
We present the resulting photometry at 3.6 and 4.5~\micron\ in
Table~\ref{tab:data}. Our new measurements at 4.5~\micron\ is fainter than
the value we reported in \citet{luh11} because of the subtraction of the
blended background source from the second epoch data.

\subsection{VLT HAWK-I}

In addition to the 3.6~\micron\ data from IRAC, we have pursued near-IR
photometry to further constrain the temperature of WD~0806-661~B 
and to assess the feasibility of spectroscopy.
We have focused on filters near 1.2-1.3~\micron\ (e.g., $J$) for these
observations since they offer the best sensitivity at near-IR wavelengths
for cool brown dwarfs. Our first set of near-IR images was obtained with the
High Acuity Wide-field K-band Imager (HAWK-I) on the Unit Telescope 4 of
the Very Large Telescope (VLT).
This camera contains four 2048$\times$2048 HAWAII-2RG arrays and has a
plate scale of $0\farcs106$~pixel$^{-1}$ \citep{kis08}.
During the nights of 2011 March 5 and 6, WD~0806-661~B was placed within
one of the four arrays and 47 dithered images were collected,
each of which consisted of 15 coadded 6.2~s exposures.
Thus, the total exposure time was 72.85~min. 
These observations were performed through program 286.C-5042.

For each of the HAWK-I images from the array containing WD~0806-661~B, 
we subtracted a dark frame and divided by a flat field image.
The resulting images were registered and combined into a single mosaic.
To measure the WCS and flux calibration for the mosaic,
we began by reducing archival $J$-band images of this field that are publicly
available from the Infrared Side Port Imager at the 4~m Blanco telescope
\citep{rod11}. 
We used astrometry and photometry from the Point Source Catalog of the 
Two-Micron All-Sky Survey \citep[2MASS,][]{skr06} to measure the flux
calibration and WCS of the ISPI mosaic, which is large enough 
($10\arcmin\times10\arcmin$) to encompass a sufficient number of 2MASS sources.
We then used ISPI astrometry and photometry to 
determine the WCS and flux calibration of the HAWK-I data.

Point sources in the HAWK-I mosaic exhibit FWHM$\sim0\farcs7$. To facilitate
visual detection of sources with low signal-to-noise ratios (SNRs) in
Figure~\ref{fig:image}, we have smoothed the image to a resolution of
$1\arcsec$. WD~0806-661~B is not detected in this image.
We estimate that SNR=3 corresponds to $J\sim23.9$.

\subsection{Magellan FourStar}

We obtained a second set of near-IR images of WD~0806-661~B using
FourStar on the Magellan 6.5~m Baade Telescope. 
This camera employs four 2048$\times$2048 HAWAII-2RG arrays and has a
plate scale of $0\farcs159$~pixel$^{-1}$ \citep{per08}.
We used a medium-band filter 
(denoted as $J3$) that extends from 1.22--1.35~\micron\ rather than a
standard $J$-band filter (1.13--1.35~\micron) since it is aligned more closely
with the wavelengths at which the near-IR fluxes of T dwarfs are highest.
During the nights of 2011 April 17 and 18, we placed WD~0806-661~B in
one of the four arrays of FourStar and collected 46 dithered images,
each of which consisted of three coadded 32~s exposures. The resulting
total exposure time was 73.6~min.

The FourStar images were reduced with procedures similar to those applied
to the HAWK-I data. We derived the flux calibration of the final mosaic
from images of two photometric standards, GD71 and GD173, and synthetic
$J-J3$ Vega-based colors derived from the transmission profiles for these
filters. To verify this calibration, we also calibrated the mosaic
using the ISPI photometry as done for HAWK-I. The ISPI sources that
appear within the FourStar data have a mean color of $J-H\sim0.5$.
Using a spectrum of Vega, we estimate that an A0V star reddened to
have this color will have $J-J3\sim0.1$. When we combined this color
with the ISPI $J$-band photometry for sources detected by FourStar,
we arrived at the same flux calibration as derived from the photometric
standards.

The average FWHM for point sources in  the FourStar mosaic is $\sim0\farcs75$.
As with the HAWK-I data, we have smoothed the FourStar image to
a resolution of $1\arcsec$ for display in Figure~\ref{fig:image}, and it
does not show a detection of WD~0806-661~B. We estimate that
SNR=3 corresponds to a Vega-based magnitude of $J3\sim23.5$ in this image.
By combining the observed spectra of late T dwarfs with the transmission
profiles for $J$ and $J3$, we derive colors of $J-J3\sim0.4$.
When we use the model spectra of brown dwarfs from \citet{bur03} and
\citet{sau08}, we arrive at similar values of $J-J3$ for temperatures above
250~K. Thus, the limit of $J3\gtrsim23.5$ implies that WD~0806-661~B has
$J\gtrsim23.9$, which is the same as the limit measured with HAWK-I.

\citet{luh11} and \citet{rod11} have previously reported $J$-band limits
for WD~0806-661~B. \citet{rod11} measured a limit of $J>21.7$ for SNR$<$3,
which they described as 1.7~mag fainter than the one from \citet{luh11}.
However, the two limits were not compared at the same SNR.
Given that the limit of $J>20$ from \citet{luh11} corresponded to SNR$>$10,
the $J$-band constraint from \citet{rod11} was 0.4~mag deeper than that
from \citet{luh11}.

\section{Analysis}
\label{sec:analysis}

\subsection{Common Proper Motion}
\label{sec:pm}

\begin{figure}
\epsscale{1.15}
\plotone{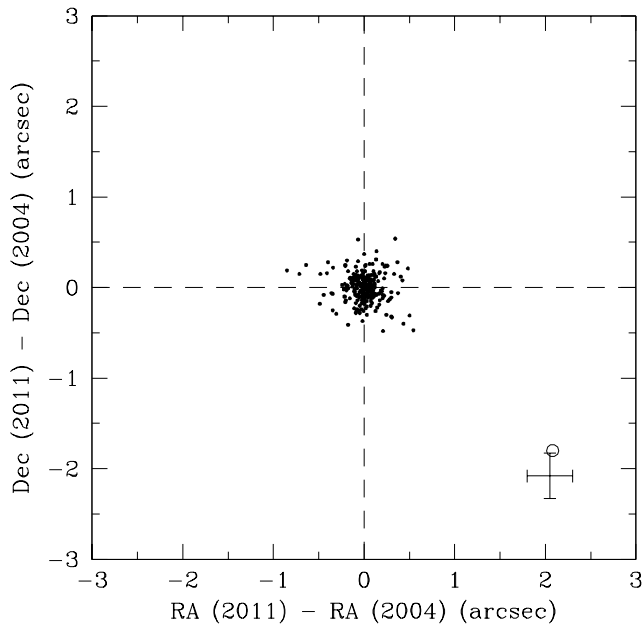}
\caption{
Differences in coordinates between the {\it Spitzer} IRAC images from 2004
and 2011 for WD~0806-661 (circle), its companion (1~$\sigma$ error bars),
and all other sources (points).
}
\label{fig:pm}
\end{figure}

\citet{luh11} identified WD~0806-661~B as a companion based on its
common proper motion with the primary between the first
two epochs of IRAC images.
Given our new detection of WD~0806-661~B at 3.6~\micron, we can verify
that WD~0806-661~B shares the same proper motion as the primary.
In Figure~\ref{fig:pm}, we show the differences in equatorial coordinates
between the 3.6~\micron\ image and the first 4.5~\micron\ image
for stars detected in both images.
To estimate the uncertainty in the displacement of the secondary, one would
normally examine the scatter in Figure~\ref{fig:pm} for sources
near its magnitude. However, because WD~0806-661~B is very red in
$[3.6]-[4.5]$, few sources in the IRAC images match its brightness at
both wavelengths. As an alternative means of estimating the uncertainties,
we have computed the standard deviations of the differences in right
ascension and declination for objects that have a similar combination
of SNRs, except in the opposite bands compared to WD~0806-661~B.
In other words, whereas WD~0806-661~B has a low SNR at 3.6~\micron\ and 
a high SNR in 4.5~\micron, we have considered sources that have low
and high SNRs at 4.5 and 3.6~\micron, respectively. Many sources of this kind
are available because the 3.6~\micron\ images are deeper than the
4.5~\micron\ data. In this way, we estimate a 1~$\sigma$ uncertainty
of $0\farcs25$ in each of the offsets in right ascension and declination.
As demonstrated in Figure~\ref{fig:pm}, the motion of WD~0806-661~B 
based on our new astrometry is non-zero at a level of 8~$\sigma$ in each
direction and is consistent within $\sim1$~$\sigma$ of that of the primary.

\subsection{Photometric Properties}

WD~0806-661~B was detected only at 4.5~\micron\ in previous images.
As a result, only limits were available for its colors, such as $J-[4.5]$,
which strongly suggested a cool temperature \citep{luh11,rod11}. 
Our detection at 3.6~\micron\ now provides a direct measurement of
a color for WD~0806-661~B. 
We find that it has $[3.6]-[4.5]=2.77\pm0.16$, which is much redder than
the neutral colors exhibited by most astronomical sources
and confirms that it is a cool, methane-bearing object \citep{pat06}.

\begin{figure}
\epsscale{1.15}
\plotone{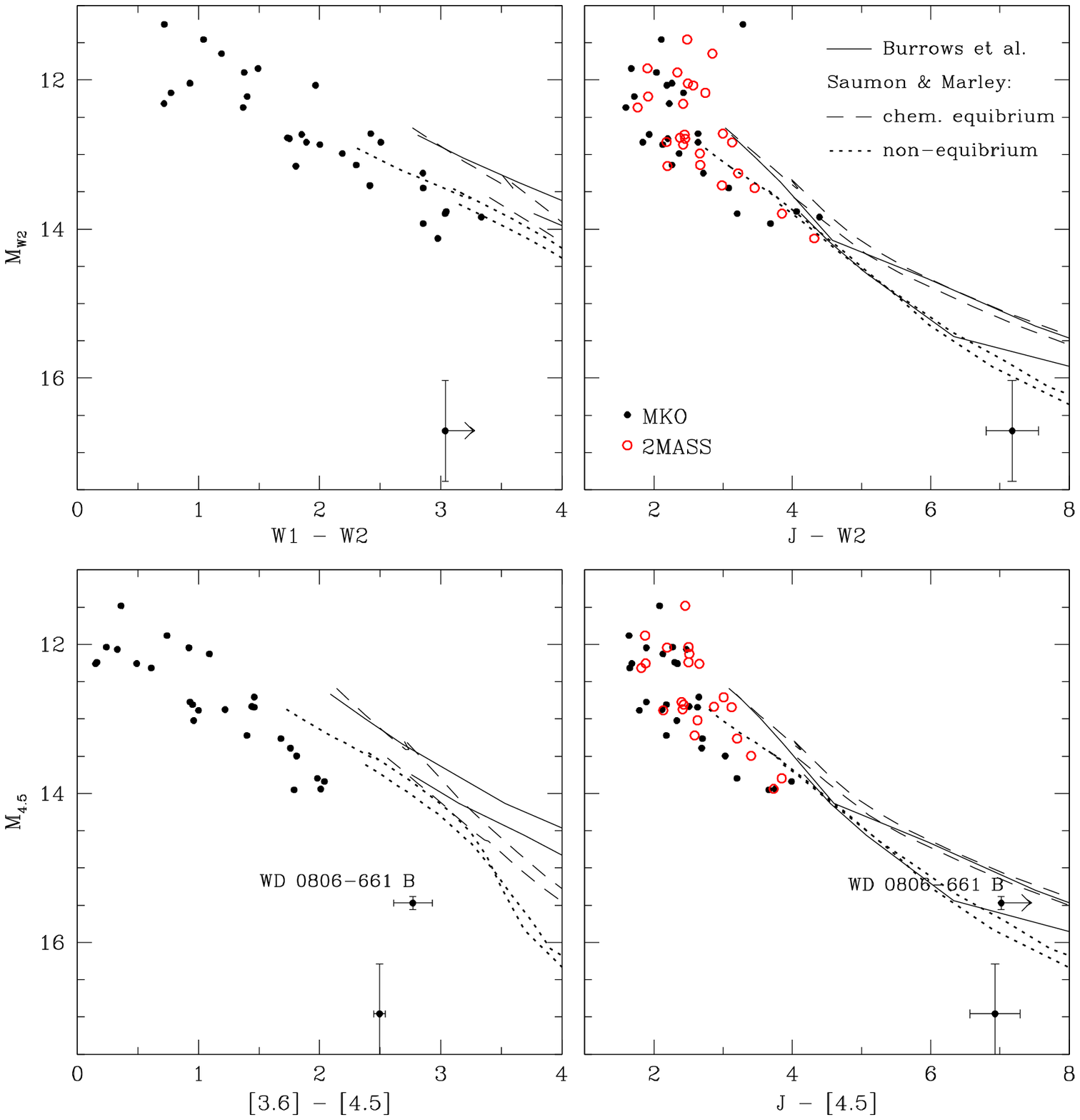}
\caption{
Color-magnitude diagrams for known T and Y dwarfs with measured distances
based on data from {\it WISE} and IRAC on {\it Spitzer}
\citep[top and bottom,][references therein]{luc10,leg10a,cus11,kir11}.
Our data for WD~0806-661~B are indicated in the bottom diagrams.
Error bars are included for WD~0806-661~B and the Y dwarf WISEP J1541-2250.
For comparison, we show the magnitudes and colors predicted for
ages of 1 and 3~Gyr by the theoretical spectra and evolutionary models
from \citet[][solid lines]{bur03} and \citet[][dashed and dotted lines]{sau08}.
}
\label{fig:cmd}
\end{figure}

To investigate how WD~0806-661~B compares to previously known brown dwarfs
in terms of temperature and luminosity, we begin by constructing
diagrams of $M_{4.5}$ versus $[3.6]-[4.5]$ and $M_{4.5}$ versus $J-[4.5]$
in Figure~~\ref{fig:cmd} for WD~0806-661~B and all known T and Y dwarfs that
have measured distances and IRAC photometry 
\citep[][references therein]{luc10,leg10a,burn11a,kir11}.
We have adopted the $J$-band photometry compiled by \citet{leg10a} and
\citet{kir11} and have plotted separately the data from the 2MASS and
Mauna Kea Observatories (MKO) systems.
Sources that have been observed in both systems are plotted twice.
When multiple $J$-band measurements are available for a given object,
we have adopted the mean of those data weighted by the inverse square of
their flux errors.  In Figure~~\ref{fig:cmd}, WD~0806-661~B is redder and
fainter than all of the T dwarfs in this sample.
It is also redder than the one Y dwarf with a measured parallax,
WISEP J154151.65-225025.2 \citep[hereafter WISEP J1541-2250,][]{cus11,kir11},
but that object appears to be fainter than WD~0806-661~B.
As a result, a single sequence of T and Y dwarfs would not be pass through
the locations of both WD~0806-661~B and WISEP J1541-2250 in
the color-magnitude diagrams, unless the former is an unresolved binary.
This discrepancy may arise from the large uncertainty in the parallax of
WISEP J1541-2250. We note that the very cool brown dwarf companion CFBDSIR
J1458+1013~B is absent from  Figure~\ref{fig:cmd} since it lacks mid-IR
photometry. It is $\gtrsim10$ times brighter than WD~0806-661~B in $M_J$, and
hence should be significantly warmer \citep[Table~\ref{tab:data},][]{liu11a}.

Since {\it WISE} has also measured mid-IR data for a large number of brown
dwarfs, we have included in Figure~\ref{fig:cmd} color-magnitude diagrams based
on the {\it WISE} bands at 3.4 and 4.6~\micron\ (denoted as $W1$ and $W2$).
These data are from \citet{mai11}, \citet{bur11}, \citet{cus11}, and
\citet{kir11}.
The photometry from the $W2$ and IRAC 4.5~\micron\ filters
have an average offset of less than a few percent for L and T dwarfs
while the $W1$ magnitudes are fainter than those from IRAC 3.6~\micron\ by an
average of $\sim$0.5 to 1.5~mag from early T dwarfs to Y dwarfs \citep{kir11}.

\begin{figure}
\epsscale{1.15}
\plotone{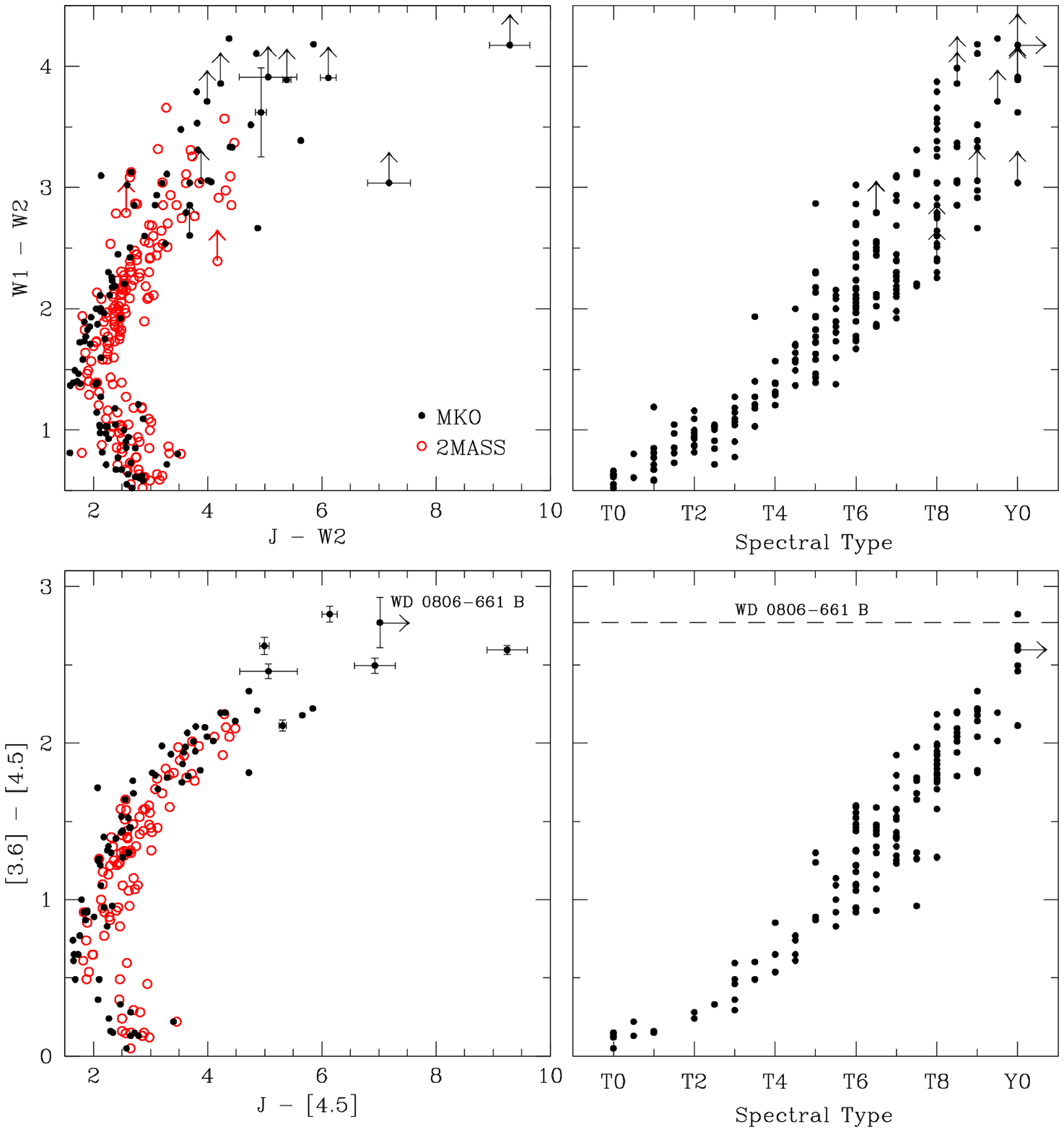}
\caption{
Color-color diagrams and colors versus spectral type for known T and Y dwarfs
based on data from {\it WISE} and IRAC on {\it Spitzer}
\citep[top and bottom,][references therein]{luc10,leg10a,burn11a,cus11,kir11}.
Our data for WD~0806-661~B are indicated in the bottom diagrams.
Error bars are included in the left diagrams for WD~0806-661~B and the Y dwarfs.
}
\label{fig:cc}
\end{figure}

In Figure~\ref{fig:cc}, we expand our comparison to include known T and Y
dwarfs that lack distance estimates by constructing diagrams that contain
only colors and spectral types. We present diagrams based on both IRAC and
{\it WISE} photometry. 
As shown in Figure~\ref{fig:cc}, WD~0806-661~B is as red as or redder than
all known T and Y dwarfs in $[3.6]-[4.5]$ and $J-[4.5]$, with the possible
exception of WISEP J1828+2650. WD~0806-661~B is marginally redder than 
that Y dwarf in $[3.6]-[4.5]$, but it is unknown how they compare in $J-[4.5]$. 
WD~0806-661~B is also redder in $[3.6]-[4.5]$ than all of the late T candidates
found in the {\it Spitzer} Deep, Wide-Field Survey \citep{eis10}.

\subsection{Physical Properties}
\label{sec:phys}

We can use our new photometry for WD~0806-661~B to update our previous
estimates of its mass and effective temperature from \citet{luh11}.
To derive the mass of WD~0806-661~B, \citet{luh11} compared $M_{4.5}$
to the predictions of evolutionary models for the age of the primary.
In that study, our age estimate was in part based on an analytic relation
between initial stellar mass and evolutionary time
through the end of core helium burning \citep{iben89}.
We now revise our previous calculation to use the evolutionary time 
predicted by the numerical models from \citet{gir02}, which results
in a new estimate of $2\pm0.5$~Gyr for the total age of the white dwarf.
When we compare our new measurement of $M_{4.5}$ to the values predicted
by \citet{bur03} and \citet{sau08} for this age, we arrive at a mass
of 6--9~$M_{\rm Jup}$ for WD~0806-661~B.

Both the colors and absolute magnitudes of cool brown dwarfs are sensitive
to their temperatures. To assess which of these measurements for
WD~0806-661~B is most likely to produce a reliable temperature estimate,
we have plotted theoretical colors and magnitudes of brown dwarfs 
in Figure~\ref{fig:cmd} for ages of 1 and 3~Gyr.
We have considered the cloudy models from \citet{bur03}, which
employ equilibrium chemistry, and new versions of the cloudless models
from \citet{sau08} for both equilibrium and non-equilibrium chemistry.
The new models based on \citet{sau08} include updates to the ammonia and
hydrogen collision induced absorption opacities \citep{yur11,fro10} and an
extension to lower effective temperatures. The non-equilibrium models are
computed with an eddy diffusion coefficient of $K_{zz}=10^4$~cm$^2$~s$^{-1}$.
As shown in Figure~\ref{fig:cmd}, all of the models are much redder than known
T dwarfs in $[3.6]-[4.5]$ and the equivalent color from {\it WISE}.
\citet{leg09,leg10a} found a similar discrepancy in the values of
$[3.6]-[4.5]$ predicted by \citet{sau08}, which they attributed to errors
in the model fluxes at 3.6~\micron.
The models agree reasonably well with the observed sequence of
T dwarfs in $M_{4.5}$ versus $J-[4.5]$. Both sets of equilibrium models are
consistent with the location of WD~0806-661~B in $M_{4.5}$ versus $J-[4.5]$ 
while the non-equilibrium calculations produce a value of $J-[4.5]$ that is
slightly too blue. Therefore, we rely on $J-[4.5]$ and $M_{4.5}$ 
for estimating the temperature of WD~0806-661~B.

\begin{figure}
\epsscale{1.15}
\plotone{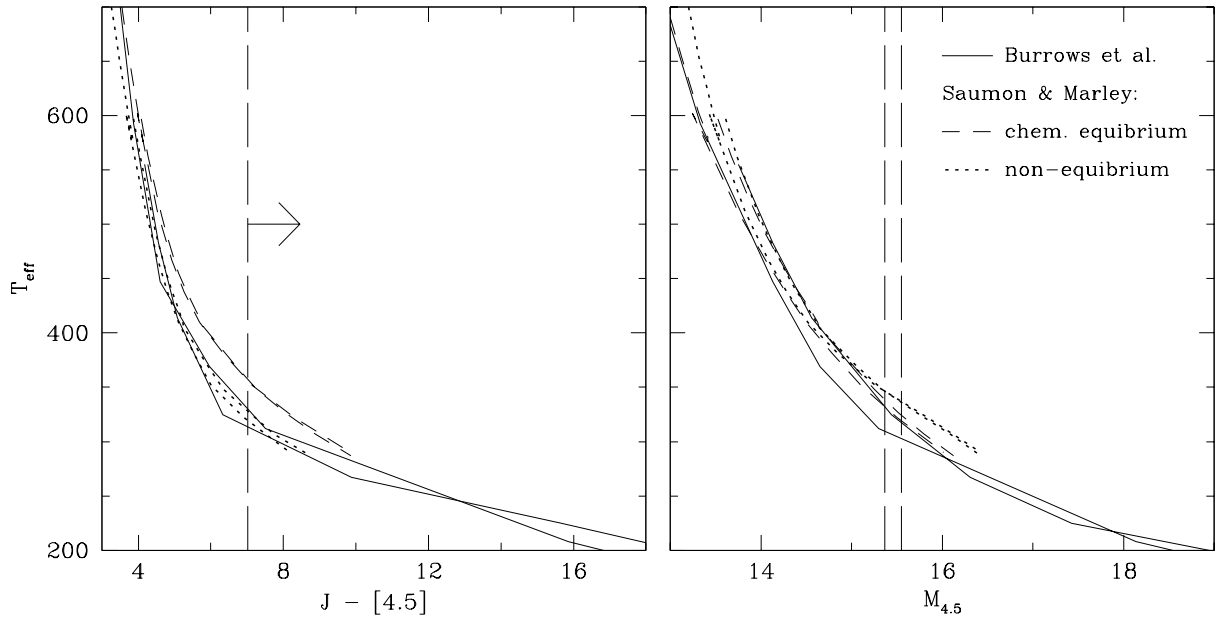}
\caption{
Predicted $J-[4.5]$ and $M_{4.5}$ as a function of effective temperature
based on the models from \citet[][solid lines]{bur03} and
\citet[][short dashed and dotted lines]{sau08} for ages of 1 and 3 Gyr.
The constraints on $J-[4.5]$ and $M_{4.5}$ for WD~0806-661~B are
indicated (long dashed lines).
}
\label{fig:teff}
\end{figure}

In Figure~\ref{fig:teff}, we compare the constraints on $J-[4.5]$ and $M_{4.5}$
to the values predicted as a function of effective temperature by the models.
The limit of $J-[4.5]>7.0$ implies $T_{\rm eff}\lesssim320$~K and
$T_{\rm eff}\lesssim355$~K for WD~0806-661~B based on the calculations
of \citet{bur03} and \citet{sau08}, respectively.
Using $M_{4.5}=15.47\pm0.09$, we derive $T_{\rm eff}=310\pm10$~K with
\citet{bur03} and $T_{\rm eff}=325\pm10$~K with \citet{sau08}.
The models for these temperature estimates assume chemical equilibrium.
The temperatures implied by $J-[4.5]$ and $M_{4.5}$ using the
non-equilibrium models are marginally inconsistent with each other
($T_{\rm eff}\lesssim325$~K, $T_{\rm eff}=340\pm5$~K), which is a reflection
of the fact that these models do not pass through the location of
WD~0806-661~B in Figure~\ref{fig:cmd}.
The quoted uncertainties represent only the error in $M_{4.5}$ and do
not include the unknown systematic errors in the model predictions.

We now examine the temperature of WD~0806-661~B in the context of
one of the faintest known brown dwarfs, CFBDSIR J1458+1013~B.
\citet{liu11a} considered two methods for estimating the temperature
of this object. In the first one, they found $T_{\rm eff}\sim400$~K for
CFBDSIR J1458+1013~B by comparing $M_J$ to the predictions of \citet{bur03},
which is similar to our approach for WD~0806-661~B.
\citet{liu11a} also compared the bolometric luminosity to the values from
evolutionary models, which produced a lower temperature of $\sim350$--380~K.
To estimate the luminosity of CFBDSIR J1458+1013~B, they assumed that
it has the same $J$-band bolometric correction (BC$_J$) as T dwarfs at
500--600~K.  However, BC$_J$ is expected to change by $\sim1.4$~mag
between 500--600~K and 350-400~K according to the models of \citet{bur03}
and \citet{sau08}.
The rapid increase in $J-[4.5]$ that is observed at lower temperatures
(Figure~\ref{fig:cc}) also indicates that BC$_J$ is unlikely to remain
constant. As a result, the bolometric luminosity of CFBDSIR J1458+1013~B
was probably underestimated, which would explain why the luminosity-based
temperature was lower than the value derived from $M_J$.

\section{Discussion}
\label{sec:disc}

Through deep imaging at $J$ and 3.6~\micron, we have further demonstrated
the companionship of WD~0806-661~B via its common proper motion with its
primary and have provided new constraints on its photometric properties.
We have found that WD~0806-661~B is redder than at least five of the six
Y dwarfs recently discovered by \citet{cus11}, indicating that it is probably
a member of this new spectral class as well.
WD~0806-661~B and the Y dwarf WISEP J1828+2650 appear to be the coldest known
brown dwarfs based on their $[3.6]-[4.5]$ and $J-[4.5]$ colors. 
A $J$-band detection of WD~0806-661~B and a parallax measurement for
WISEP J1828+2650 are needed to determine their relative temperatures.
More accurate photometry at 3.6~\micron\ for the former and at $J$ for the
latter also would help refine the comparison of these objects.

In addition to representing one of the coldest known objects directly observed
outside the solar system, WD~0806-661~B is a benchmark for
testing atmospheric and evolutionary models at substellar masses
since its age and distance are well-constrained via its primary.
To fully exploit it for this purpose, deeper imaging at $J$
and other bands is necessary. WD~0806-661~B is too faint for ground-based
spectroscopy, but near-IR grism observations with the
{\it Hubble Space Telescope} may be feasible.
The new Y dwarfs found with {\it WISE} have closer distances and
are much more amenable to detailed spectroscopic study \citep{cus11},
but even low-SNR spectroscopy of WD~0806-661~B would be valuable given
its extreme temperature and benchmark status.

Because of the rapid decrease in the $J$-band fluxes of brown dwarfs
at the lowest temperatures, it seems unlikely that objects cooler than
WD~0806-661~B will be found with wide-field ground-based imaging, which is
restricted to near-IR wavelengths. For instance, current ground-based surveys
have identified T dwarfs down to $J\sim19.5$ \citep{burn10,del10},
corresponding to a distance limit of $<$2.5~pc for analogs of WD~0806-661~B.
The discovery of any additional objects with $T_{\rm eff}\lesssim350$~K
over the next decade will probably occur through continued analysis of
mid-IR data from the {\it WISE} and {\it Spitzer} satellites.
For instance, typical mid-IR images from {\it WISE} are capable of
detecting objects like WD~0806-661~B with SNR=10 out to distances of
$\sim7$~pc. It also is possible that companions in this temperature regime
could be uncovered through very deep near-IR imaging of the nearest stars with
ground-based adaptive optics or the {\it Hubble Space Telescope}.

\acknowledgements

We acknowledge support from grant AST-0544588 from the National Science
Foundation (K. L., J. B.) and the NASA Astrophysics Theory Program
(M. M., D. S.).
This publication makes use of data products from the following resources:
the {\it Wide-field Infrared Survey Explorer}, which is a joint project of the
University of California, Los Angeles, and the Jet Propulsion
Laboratory/California Institute of Technology, funded by the National
Aeronautics and Space Administration;
the NASA/IPAC Infrared Science Archive, which is operated by the Jet
Propulsion Laboratory, California Institute of Technology, under contract
with the National Aeronautics and Space Administration;
the SpeX Prism Spectral Libraries, maintained by Adam Burgasser at
http://www.browndwarfs.org/spexprism;
the M, L, and T dwarf compendium housed at http://DwarfArchives.org and
maintained by Chris Gelino, Davy Kirkpatrick, and Adam Burgasser.
The Center for Exoplanets and Habitable
Worlds is supported by the Pennsylvania State University, the Eberly College
of Science, and the Pennsylvania Space Grant Consortium.

\begin{deluxetable}{ll}
\tabletypesize{\scriptsize}
\tablewidth{0pt}
\tablecaption{Properties of WD~0806-661~B\label{tab:data}}
\tablehead{
\colhead{Parameter} & \colhead{Value} \\
}
\startdata
Distance & 19.2$\pm$0.6 pc\tablenotemark{a} \\
Age & 2$\pm$0.5 Gyr\tablenotemark{b} \\
$J$ & $>$23.9\tablenotemark{c} \\
$J3$ & $>$23.5\tablenotemark{c,d} \\
$[3.6]$ & 19.65$\pm$0.15 \\
$[4.5]$ (2004) & 16.96$\pm$0.09 \\
$[4.5]$ (2009) & 16.84$\pm$0.06 \\
$[4.5]$ (mean) & 16.88$\pm$0.05 \\
$M_{4.5}$ & 15.47$\pm$0.09 \\
$T_{\rm eff}$ & 300--345~K\tablenotemark{e} \\
Mass & 6--9~$M_{\rm Jup}$\tablenotemark{e} \\
\enddata
\tablenotetext{a}{Measured for the primary \citep{sub09}.}
\tablenotetext{b}{Estimated for the primary (Section~\ref{sec:phys}).}
\tablenotetext{c}{SNR$<$3.}
\tablenotetext{d}{WD~0806-661~B likely has $J-J3\sim0.4$.}
\tablenotetext{e}{Based on a comparison of $M_J$ and $J-[4.5]$ to the values
predicted by the models in Figure~\ref{fig:teff}.}
\end{deluxetable}

\clearpage

\end{document}